\documentclass[pra,twocolumn,superscriptaddress,10pt,showpacs]{revtex4-1}
\usepackage[pdftex,plainpages=false,colorlinks=true,citecolor=blue,linkcolor=blue,urlcolor=blue,filecolor=green,bookmarksopen=true]{hyperref}
\usepackage[utf8]{inputenc}
\usepackage[portuguese]{babel}
\usepackage{amsmath}
\usepackage{subfig}
\usepackage{graphicx,epstopdf}
\usepackage{amsfonts}
\usepackage{bbm}
\usepackage{amssymb}
\usepackage{enumerate}
\usepackage{color}
\usepackage{latexsym}
\usepackage{times,txfonts}

\newcommand{\1}{\mathbbm{1}}

\newcommand{\ket}{\rangle}

\begin{document}

\title{O Computador Quântico da IBM e o \textit{IBM Quantum Experience}}

\author{Alan C. Santos}
\email{alancs@if.uff.br}
\affiliation{Instituto de F\'{i}sica, Universidade Federal Fluminense, Av. Gal. Milton Tavares de Souza s/n, Gragoat\'{a}, 24210-346 Niter\'{o}i, Rio de Janeiro, Brazil}

\begin{abstract}

O anúncio de um computador quântico que pode ser acessado de forma remota por qualquer pessoa a partir de seu laptop é um acontecimento de grande importância para os cientistas da computação quântica. Neste trabalho nós apresentamos o computador quântico da \textit{International Business Machines} (IBM) e sua plataforma \textit{IBM Quantum Experience} (IBM-QE) como uma proposta didática em estudos de computação e informação quântica e, não menos importante, como divulgação científica de tal anúncio feito pela equipe da IBM. Apresentamos as principais ferramentas (portas quânticas) disponíveis no IBM-QE e, por meio de uma simples estratégia, discutimos acerca de uma das fontes de decoerência nos chips de 5 q-bits da IBM. Como exemplo de aplicação do nosso estudo, nós mostramos como implementar o teleporte quântico usando o IBM-QE. 

\end{abstract}

\maketitle

\section{Introdução}

Assim como o conceito de teleporte quântico, tecnologias quânticas hoje não são mais apenas elementos principais de filmes de ficção científica. Devido ao grande avanço de estudos em mecânica quântica e a tentativa de entender a natureza quântica do universo e como usá-la em algo útil, hoje vivemos um cenário que antes eram apenas possíveis na imaginação. Na primeira metade da década de 80 iniciava-se a elaboração dos fundamentos que sustentam a pesquisa em computação quântica (CQ), graças aos trabalhos de Paul Benioff \cite{Benioff:80,Benioff:82}, Richard Feymman \cite{Feynman:82} e David Deutsch \cite{Deutsch:85}. A partir desses trabalhos, o sonho do computador qu\^{a}ntico tem sido buscado devido a sua capacidade te\'{o}rica de resolver certas classes de problemas muito mais
r\'{a}pido que um computador cl\'{a}ssico.

No dia de 04 de maio de 2016, a IBM e seus cientistas da computação quântica anunciaram aquela que provavelmente foi a mais motivadora notícia para muitos que fazem pesquisas nessa área. O primeiro computador quântico de acesso remoto e publico foi disponibilizado pela equipe da IBM. A ideia é que qualquer pessoa (não necessariamente pesquisadores da área) possam ter acesso remoto a uma plataforma conhecida como \textit{IBM Quantum Experience} (IBM-QE) \cite{IBM}. A proposta da IBM é que com essa plataforma qualquer pessoa possa simular e até mesmo executar uma computação em um computador quântico de $5$ q-bits. É claro que não se pode fazer muito com apenas $5$ q-bits, mas a ideia é que possamos testar o IBM-QE para determinar se realmente estamos lidando com um computador quântico de fato. Nesse intuito, o IBM-QE tem sido usado recentemente em diversos protocolos de CQ \cite{Latorre:16,Devitt:16,Berta:16,Rundle:16}, em especial tem-se ilustrado o experimento real por meio do teleporte quântico onde uma discussão mais detalhada do aparato experimental é feito \cite{Fedortchenko:16}. Essa determinação é o atual problema que cientistas da CQ estão tentando resolver em protótipos como os chips quânticos da IBM, ditos quânticos. Neste paper nós apresentamos a comunidade as características e sutilezas do IBM-QE, que tem duas propostas básicas. A primeira delas é a divulgação científica de tal aparato, a segunda é uma proposta didática de que o IBM-QE possa ser uma ferramenta com certo valor ao introduzirmos certos conceitos em computação e informação quântica. O paper está estruturado da seguite forma.

Na seção \ref{Elementos} nós faremos uma breve introdução à CQ, mas sempre focando em discutir algumas definições que serão úteis para o nosso desenvolvimento. No entanto, quando necessário deixaremos referências de livros-texto da área onde informações adicionais podem ser encontradas. Na subseção \ref{QuantumTeleporte} nós apresentaremos o modelo de teleporte quântico e como podemos implementá-la em um computador quântico. Na seção \ref{IBMQE} nós apresentaremos o computador quântico da IBM e o IBM-QE, discutindo suas capacidades e características, onde mostramos um protocolo que pode ser usado para analisar os efeitos de decoerência no computador quântico da IBM, que será discutido na Seção \ref{Deco}. Como exemplo, na subseção \ref{QuantumTeleporteExpe}, nós implementamos o protocolo de teleporte e discutimos como interpretar os resultados fornecidos pela plataforma IBM-QE.

\section{Elementos de computação quântica} \label{Elementos}

Antes de discutir as propriedades e serviços disponíveis no computador quântico da IBM, bem como na sua plataforma IBM-QE, deixe-nos dedicar essa seção a discutir os fundamentos da computação quântica. Para uma abordagem mais detalhada, recomendamos uma boa referência como sendo a Ref. \cite{Nielsen:Book}

\subsection{Bits quâticos (q-bits) e Emaranhamento}

Em computadores clássicos, a menor unidade de informação é definida como \textit{bit} e, a partir deste, podemos definir outras quantidades como o \textit{byte} (8 bits), \textit{megabyte} ($10^6$ bytes), e assim por diante. De forma análoga, também temos uma unidade fundamental de informação em computação quântica e, não por coincidência, chamamos de \textit{q-bit}, que é a tradução para português da definição \textit{qu-bit} que, por sua vez, é uma forma compacta para nos referirmos a um \textit{quantum bit} (bit quântico). Mas nos perguntamos: O que ganhamos de interessante (novo) nessa história toda?

Em geral, os \textit{q}-bits são representados fisicamente por estados ortogonais associados a qualquer sistema quântico de dois níveis de energia (estados de polarização vertical e horizontal de fótons, estados de spin do elétron, etc.) e denotados pelos \textit{vetores de estado} abstratos $\vert 0 \ket$ e $\vert 1 \ket$, que são os chamados \textit{estados da base computacional} (em analogia ao $0$ e $1$ de computadores clássicos). O que ganhamos na transição do bit para o \textit{q}-bit é que existem peculiaridades da mecânica quântica que nos permite combinar esses estados a fim de obter alguma vantagem com relação aos bits. De fato, considere o estado de superposição $\vert \psi \ket = a \vert 0 \ket + b\vert 1 \ket$, onde $\vert a \vert^2+\vert b \vert^2=1$ (condição de normalização do estado $\vert \psi \ket$), que é um tipo de configuração muito comum em mecânica quântica. Podemos perceber que o estado $\vert \psi \ket$ representa uma superposição de estados da base computacional, ao mesmo tempo que nós precisamos apenas de uma única partícula para "escrever" esse estado, o que não é possível em computadores clássicos. A vantagem do computador quântico aparece quando consideramos mais de um \textit{q}-bit. Deixe-nos considerar um exemplo muito simples de um  estado particular de um sistema composto por dois sistemas de dois níveis $\vert \psi_2 \ket = \frac{1}{2}(\vert 00 \ket+\vert 01 \ket+\vert 10 \ket+\vert 11 \ket)$, que carrega todas as combinações possíveis para dois bits. Podemos perceber que com apenas dois \textit{q}-bits nós podemos armazenar uma quantidade de informação que requer 8 bits (1 byte) em computadores clássicos, isso nos dá uma noção de que podemos ganhar uma economia muito boa em espaço físico com tecnologias quânticas.

Outro recurso fundamental em computadores quânticos, bem como em toda a teoria da informação quântica, é o chamado \textit{emaranhamento}. De forma simples, podemos definir o emaranhamento como uma quantidade física (pode ser mensurada) associada a duas partículas que nos impossibilita de caracterizar completamente o estado de uma partícula independente da segunda. Existe também uma visão matemática que é mais simples de caracterizar o emaranhamento, que é a noção de separabilidade. Para exemplificar, considere o seguinte estados de dois \textit{q}-bits $A$ e $B$ como
\begin{equation}
\vert \psi_2 \ket_{\text{AB}} = a\vert 00 \ket_{\text{AB}} + b\vert 01 \ket_{\text{AB}} + c\vert 10 \ket_{\text{AB}} + d\vert 11 \ket_{\text{AB}}  \text{ \ ,} \label{state2qubits}
\end{equation}
então dizemos que este estado é separável (não emaranhado) se, e somente se existem coeficientes $\alpha , \beta , \gamma, \delta$ tais que podemos escrever o estado acima como\begin{equation}
\vert \psi_2 \ket_{\text{AB}} = (\alpha \vert 0 \ket + \beta\vert 1 \ket)_{\text{A}}( \gamma \vert 0 \ket + \delta \vert 1 \ket)_{\text{B}} \text{ \ .}
\end{equation}

Embora essa forma como definimos acima o emaranhamento como consequência da não separabilidade seja independente de qual sistema estamos interessados, existem situações onde o emaranhamento não é tão simples de perceber. Em geral, quando temos estados mistos a situação pode ser mais drástica e assim recomendamos a Ref. \cite{Nielsen:Book} para um estudo mais aprofundado sobre o tema. Para os exemplos que vamos tratar nesse material, o método acima é o mais indicado, dado sua simplicidade.

O emaranhamento desempenha um papel essencial em teoria da informação e computação quântica, pois quando combinado com a propriedade de superposição, este nos fornece protocolos que ilustram grande eficiência de um computador quântico com relação a um computador clássico. Uma esmagadora parte dos pesquisadores em computação e informação quântica destacam o algoritmo de Shor \cite{Shor:94} como sendo o exemplo mais claro da vantagem de um computador quântico com relação a um computador clássico. O algoritmo de Shor é usado para fatorar eficientemente números primos com muitos dígitos que são intratáveis em computadores clássicos. Além desse, podemos mencionar o algoritmo de Deusch-Josza \cite{Jozsa:92} (diferenciar funções constantes de funções balanceadas) e algoritmo de Grover \cite{Grover:96,Grover:97} (algoritmo de busca em uma lista desordenada) como algoritmos que ilustram o teórico potencial esperado por um computador clássico. Essa discussão feita é muito introdutória, uma vez que o nosso foco não é falar da distinção entre computadores clássicos e quânticos, mas deixamos como uma fonte de leitura (uma boa leitura, por sinal) a Ref. \cite{Ernesto:Book}, onde o autor destaca de forma clara o surgimento da computação quântica e do seu impacto no desenvolvimento tecnológico.

\subsection{Portas e circuitos quânticos}

O modelo padrão de CQ é chamado de \textit{Modelo de Circuitos} \cite{Barenco:95}. Nos propomos a discutir sobre esse modelo pelo fato de que é esse modelo de computação que está disponível para o usuário no IBM-QE. A ideia do modelo de circuitos em computação quântica é equivalente ao que é usado em computação clássica, onde representamos o processo de computação através de uma sequência de portas lógicas quânticas (transformações unitárias em mecânica quântica) que são aplicadas a uma configuração de entrada (input) e nos fornece uma configuração de saída com o resultado da computação (output). A motivação desse modelo repousa sobre o fato de que existem manipulações que podemos fazer sobre o estado quântico de uma partícula que nos faz lembrar da ação de portas em computadores clássicos. Por esse motivo nós definimos as \textit{portas quânticas} de um \textit{circuito quântico}, que juntos são os análogos quânticos das portas que compõe um circuito em um computador clássico. 

Um exemplo de porta quântica que tem um análogo em computação clássica é a porta NOT, cuja ação inverte o bit de entrada de $0 \rightarrow 1$ ou $1 \rightarrow 0$, que tem como análogo quântico a porta quântica representada pelo operador de Pauli $\sigma_x$, que ao atuar em um estado de spin$-\frac{1}{2}$ na direção $z$ inverte o estado de $\vert +\frac{1}{2} \ket \rightarrow \vert -\frac{1}{2} \ket$ e de $\vert -\frac{1}{2} \ket \rightarrow \vert +\frac{1}{2} \ket$. Por outro lado, existem portas exclusivas da CQ, como a porta Hadamard, onde sua atuação leva estados da base computacional em superposições desses estados e vice-versa. Uma característica comum entre CQ e computação clássica são os chamados \textit{conjuntos de portas universais para computação} \cite{Barenco:95}. Esses conjuntos são assim chamados devido ao fato de seus elementos poderem ser combinados para realizar qualquer porta lógica de um circuito.

Assim como em computa\c{c}\~{a}o cl\'{a}ssica, n\'{o}s temos um conjunto de
portas elementares em CQ \cite{Nielsen:Book}.
Dentro desse conjunto de portas elementares, n\'{o}s podemos identificar
subconjuntos de portas que podem ser usados para construir os chamados 
\textit{conjuntos universais de portas qu\^{a}nticas}. Por defini\c{c}\~{a}o
esses conjuntos s\~{a}o compostos por portas elementares que podem ser combinadas de tal maneira que nos permite simular o funcionamento de qualquer porta de um circuito quântico. Exemplos desses conjuntos s\~{a}o os conjuntos \{$CNOT$ + Rota\c{c}\~{o}es de 1 q-bit\}\cite{Barenco:95} e o conjunto \{Toffoli, Hadamard\}
\cite{Kitaev:97,Dorit:03}. Além disso, existem conjuntos de portas que permitem universalidade \textit{aproximada}, como o conjunto \{$CNOT$ + Hadamard + porta $\frac{\pi }{8}$\} \cite{Boykin:00}. Quando necessário, voltaremos a discutir sobre algumas portas quânticas que serão de interesse para nosso desenvolvimento, mas novamente voltamos a recomendar uma leitura aprofundada em livros-texto da área para maiores detalhes.

\subsection{Exemplo: Circuito para o teletransporte quântico} \label{QuantumTeleporte}

O teleporte quântico (TQ) foi proposto em 1993 por Bennett \textit{et al.} \cite{Bennett:93}. Este constitui um canal para enviar informação codificada em um estado quântico desconhecido de uma parte (chamada Alice) para outra (chamada Bob) separadas espacialmente. O principal resultado do TQ é que, além de não ser necessário conhecer o estado a ser teleportado, não há qualquer limite para a distância entre os agentes (exceto pelo canal clássico que deve ser estabelecido entre eles). Por exemplo, experimentos recentes para implementação de TQ atigiram a marca dos $100$ km com fibras ópticas \cite{Takesue:15} e $143$ km no espaço livre \cite{Ma:12}. 
Um dos recursos necessários para que possamos implementar o TQ é que Alice e Bob devem compartilhar de qualquer um dos estados emaranhados abaixo
\begin{equation}
\vert \beta _{nm}\ket =\frac{\vert 0n\ket
+\left( -1\right) ^{m}\vert 1\bar{n}\ket }{\sqrt{2}} \text{ \ \ ,}
\label{BellState}
\end{equation}%
que são conhecidos na literatura como \textit{estados de Bell}, onde $n,m$ assumem valores $0$ e $1$ e onde definimos $\bar{n}=1-n$. O sistema que compõe o protocolo de teleporte é composto essencialmente por 3 \textit{q}-bits, onde dois deles estão em posse da Alice e um terceiro em posse do Bob. O sistema da Alice é composto por uma partícula do par emaranhado dado na Eq. (\ref{BellState}), onde a outra está em posse do Bob, e uma partícula onde deve estar codificada a informação a ser enviada por Alice. Considerando que o estado a ser eviada pela Alice é escrito como $\vert \psi \ket =a\vert 0\ket +b\vert 1\ket $, então o estado inicial do sistema fica escrito como
\begin{equation}
\vert \phi \ket  =\vert \psi \ket_{A} \vert \beta _{11}\ket_{AB} =\left( a\vert 0\ket +b\vert 1\ket \right)_{A} \left( \frac{\vert 01\ket -\vert 10\ket }{\sqrt{2}}\right)_{AB} \label{InicialEstado}
\end{equation}
onde em particular adotamos o canal quântico entre Alice e Bob (par emaranhado) como sendo o estado dado na Eq. (\ref{BellState}) com $(n,m)=(1,1)$. 

Devido a separa\c{c}\~{a}o entre Alice e Bob, Alice n\~{a}o pode
realizar nenhum tipo de medida na part\'{\i}cula do Bob, mas existe uma
medida especial que Alice pode fazer em suas part\'{\i}culas que possibilita
o TQ do estado $\vert \psi \ket $ para Bob. Em suas
particulas Alice deve realizar uma medida na base de Bell, um dos estados (\ref{BellState}). Assim, \'{e} conveniente escrever o
estado do sistema numa base onde as part\'{\i}culas da Alice estejam
escritas na base de Bell. Fazendo isso n\'{o}s encontramos%
\begin{eqnarray}
\vert \phi \ket  &=&\frac{1}{2}\left[ \vert \beta
_{11}\ket _{A}\left( -a\vert 0\ket -b\vert
1\ket \right) _{B}+\vert \beta _{10}\ket _{A}\left(
-a\vert 0\ket +b\vert 1\ket \right) _{B}\right] \notag
\\
&&+\frac{1}{2}\left[ \vert \beta _{01}\ket _{A}\left(
a\vert 1\ket +b\vert 0\ket \right)
_{B}+\vert \beta _{00}\ket _{A}\left( a\vert
1\ket -b\vert 0\ket \right) _{B}\right] \text{ \ \ ,}
\end{eqnarray}%
onde fica claro os poss\'{\i}veis resultados da Alice quando ela realizar
uma medida na base $\vert \beta _{nm}\ket $. Tamb\'{e}m \'{e} 
\'{o}bvio que o estado, para o qual a part\'{\i}cula do Bob ir\'{a} colapsar
depois da medida realizada pela Alice, depende exclusivamente do resultado
da medida da Alice. Exceto no caso onde o resultado da Alice fornece o
estado $\vert \beta _{11}\ket $ em suas part\'{\i}culas,
qualquer outro resultado n\~{a}o pode caracterizar o TQ, pois os poss%
\'{\i}veis estados de colapso s\~{a}o diferentes do estado original $%
\vert \psi \ket $.

\begin{table}
\centering
\caption{Correções do Bob para o TQ de um estado desconhecido}
\begin{tabular}{c|c}
\hline
Resultado da Alice & Corre\c{c}\~{a}o do Bob \\ \hline
$\vert \beta _{00}\ket $ & $\sigma _{z}\sigma _{x}$ \\ 
$\vert \beta _{10}\ket $ & $\sigma _{z}$ \\ 
$\vert \beta _{01}\ket $ & $\sigma _{x}$ \\ 
$\vert \beta _{11}\ket $ & $\1$ \\ \hline
\end{tabular}
\label{TabelaTQSimples}
\end{table}

Eis agora que surge a necessidade e a import\^{a}ncia de possibilitar a troca de informa\c{c}\~{a}o entre
Alice e Bob por meio de um canal cl\'{a}ssico. Quando falamos aqui canal clássico, estamos nos referindo a algum canal que possibilite a troca de bits de informação entre Alice e Bob, isto é, qualquer meio de comunicação no qual a velocidade de propagação da informação por esse canal esteja vinculada a velocidade da luz. Note que, para o Bob, depois
de uma medida da Alice o estado n\~{a}o est\'{a} bem definido e pode, ou n%
\~{a}o, ser o estado que Alice desejava teleportar. Mas se a Alice informar
o resultado de sua medida para o Bob, este sempre poder\'{a} agir sobre sua
part\'{\i}cula e "recuperar" a informa\c{c}\~{a}o que ficou embaralhada com
o TQ. A Tabela \ref{TabelaTQSimples} mostra as corre\c{c}\~{o}es que devem ser feitas
por Bob em sua part\'{\i}cula.

\begin{figure}[!htb]
\centering
\includegraphics[scale=0.3]{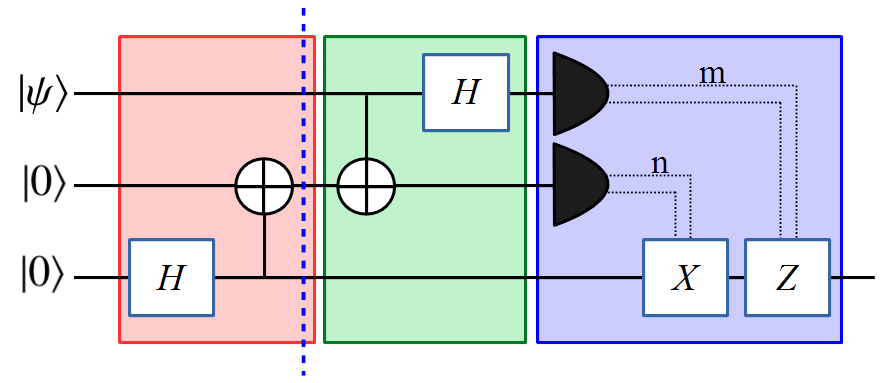}
\caption{Circuito quântico para a implementação do teleporte quântico.}
\label{BBCprotocol}
\end{figure}

Em particular, nosso interesse no protocolo do TQ é que nós podemos construir um circuito quântico capaz de implementá-lo, como mostrado por Gilles Brassard, Samuel L. Braunstein e Richard Cleve em 1996 \cite{Brassard:96} (depois estudado e aprimorado por Daniel Gottesman e Isaac Chuang \cite{Gottesman:99}). No protocolo Brassard-Braunstein-Cleve (BBC) o circuito que implementa o teleporte deve levar em conta que o estado inicial do sistema deve estar inicialmente no estado $\vert 000 \ket$ e assim devemos implementar o seguinte circuito ilustrado na Fig. \ref{BBCprotocol}.

O circuito do protocolo BBC é composta de três subcircuitos que realizam operações diferentes, mas que para realizar o teleporte eles são importantes. 

\begin{itemize}

\item Subcircuito I (vermelho): Nessa parte do circuito, nós preparamos o estado emaranhado $\vert \beta _{00}\ket $ entre os \textit{q}-bits 2 e 3. Embora o estado emaranhado seja preparado nesses \textit{q}-bits, isso não é uma regra e devemos levar em conta que nesse caso o estado a ser teleportado deve estar codificada no \textit{q}-bit 1.

\item Subcircuito II (verde): Em computadores clássicos nós podemos apenas ler "0" ou "1", mas isso não é verdade em computadores quânticos onde nós podemos ler desde "0" ou "1" bem como superposições deles. No entanto, experimentalmente é mais conveniente fazermos leituras de "0" ou "1" em computadores quânticos, o que equivale a fazer medidas na base computacional $\vert 0 \ket$ e $\vert 1 \ket$. O segundo trecho do circuito nos permite simular uma medida de Bell (feita originalmente no primeiro protocolo de TQ), mas efetivamente a medida é feita na base $\vert 0 \ket$ e $\vert 1 \ket$. Assim, desde que esse trecho do circuito seja implementado, não precisamos fazer uma medida na base de Bell e fazemos a medida na base $\vert 0 \ket$ e $\vert 1 \ket$.

\item Subcircuito III (azul): Assim como no modelo original do TQ, após a ação dos circuitos I e II o estado do sistema terá colapsado (após a medida) para um estado tal que o TQ não, necessariamente, está configurado. Assim, novamente precisamos de correções que dependem do resultado da medida na base computacional. Adotando que o reultado da medida seja $(m,n)= \{ (0,0),(0,1),(1,0),(1,1) \}$, então temos uma correção associada a cada par $(m,n)$. Isso é devido ao fato de que antes da medida o sistema estará no estado computado $\vert \phi _{C}\ket $ dado por
\begin{eqnarray}
\vert \phi _{C}\ket  &=&\frac{1}{2}\left[ \vert
00\ket _{A}\left( a\vert 0\ket +b\vert
1\ket \right) _{B}+\vert 01\ket _{A}\left(
a\vert 1\ket +b\vert 0\ket \right) _{B}\right] 
\notag \\
&&+\frac{1}{2}\left[ \vert 10\ket _{A}\left( a\vert
0\ket -b\vert 1\ket \right) _{B}+\vert
11\ket _{A}\left( a\vert 1\ket -b\vert
0\ket \right) _{B}\right] \text{ .} \label{ComputatedState}
\end{eqnarray}

As corre\c{c}\~{o}es necessárias para os respectivos resultados da medida s\~{a}o
mostradas na Tabela \ref{TabelaCircuitoTQSimples}. Anteriormente usamos um outro estado emaranhado para implementar o teleporte, mostrando assim que o teleporte acontece independente do estado escolhido para implementar o teleporte. No entanto, como pode ser visto nas Tabelas \ref{TabelaCircuitoTQSimples} e \ref{TabelaTQSimples}, as correções mudam quando mudamos o estado emaranhado usado como recurso, isso nos permite uma certa segurança ao usar o TQ para troca de informação sigilosa.

\end{itemize}

\begin{table}
\centering
\caption{Correções para o TQ via circuitos}
\begin{tabular}{c|c}
\hline
Resultado da medida & Corre\c{c}\~{a}o \\ \hline
$\vert 00\ket $  & $\1$ \\ 
$\vert 01\ket $  & $\sigma _{x}$ \\ 
$\vert 10\ket $  & $\sigma _{z}$ \\ 
$\vert 11\ket $  & $\sigma _{z}\sigma _{x}$
\\ \hline
\end{tabular}
\label{TabelaCircuitoTQSimples}
\end{table}

É importante mencionar que essa exigência feita no protocolo BBC leva em conta que todos os \textit{q}-bits devem estar próximos um dos outros, e isso não é conveniente se desejamos enviar informação de um lugar a outro. Isso é evidente, uma vez que não precisamos de um telefone para falar com alguém sentado ao nosso lado. No entanto, a proposta do protocolo BBC é que nós podemos usar o TQ para nos auxiliar em uma computação em computadores quânticos, e nesse cenário o protocolo BBC é conveniente. Se desejamos usar o protocolo BBC para enviar comunicação entre duas partes espacialmente separadas, nós podemos considerar o caso onde o estado inicial do sistema não é $\vert 000 \ket$, mas sim, como dado na Eq. (\ref{InicialEstado}). Dessa forma nós reduzimos nosso circuito de modo que usamos um circuito como no protocolo BBC, mas apenas a partir da linha pontilhada.

\section{O \textit{IBM Quantum Experience} (IBM-QE)} \label{IBMQE}

Nesta seção nós pretendemos apresentar a plataforma IBM-QE e discutir sobre algumas propriedades e limitações de tal plataforma. Vamos nos restringir a detalhar apenas aqueles componentes que serão de utilidade para nosso desenvolvimento e implementação do protocolo de teleporte, deixando que o leitor interessado em mais detalhes da plataforma IBM-QE possa usufruir de tal serviço pelo link da Ref. \cite{IBM}.

\subsection{O Computador Quântico da IBM}

\begin{figure}[!htb]
\centering
\subfloat[Chip da IBM]{
\includegraphics[scale=0.5]{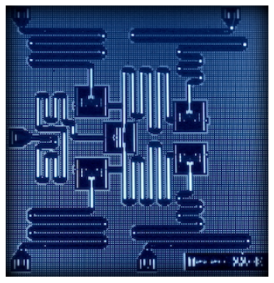}
\label{chip}
}
\quad 
\subfloat[Localização dos q-bits no Chip]{
\includegraphics[scale=0.5]{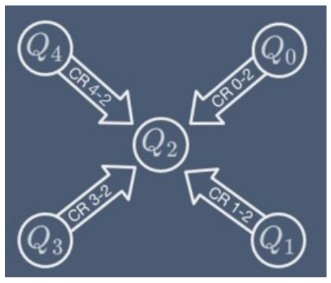}
\label{bitschip}
}
\quad 
\subfloat[Interface do IBM-QE \cite{IBM}]{
\includegraphics[scale=0.36]{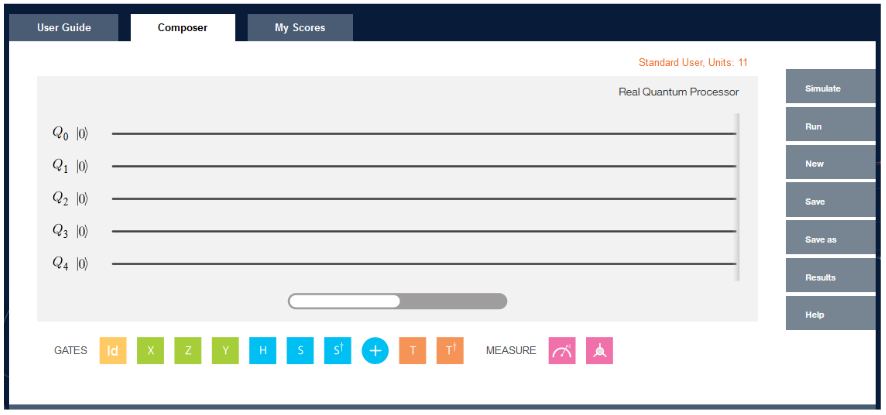}
\label{circuito}
}
\caption{(\ref{chip}) Visualização de um chip de 5-qbtis da IBM que pode ser encontrado em \cite{Link1}. (\ref{bitschip}) Esquema da disposição dos q-bits em um chip da IBM, onde a numeração está de acordo com (\ref{circuito}) interface digital da plataforma IBM-QE para um processador real}.
\label{circuitos}
\end{figure}

O computador quântico da IBM é um aparato experimental que baseia-se no modelo de computação quântica via  fluxo magnético em q-bits supercondutores \cite{Clarke:08}. O grupo IBM nos fornece, em sua plataforma \cite{IBM}, a possibilidade de trabalharmos com um chip contendo $5$ q-bits onde podemos manipulá-los a nosso critério com a finalidade de obtermos o resultado de uma dada computação. Na Fig. \ref{circuitos} nós podemos ver um esquema de como os q-bits estão dispostos em um chip da IBM e da plataforma do IBM-QE. A Fig. \ref{circuito} apresenta basicamente a área de trabalho do IBM-QE, onde seus componentes e suas propriedades serão o foco de nossa discussão agora. 

Na área de trabalho as informações sobre como trabalhar com o IBM-QE, bem como algumas revisões de elementos de computação quântica, podem ser obtidos na aba \textit{User Guide}. A aba \textit{My Scores} é onde estarão armazenados todas os circuitos construídos pelo usuário, de modo que estes não necessariamente precisam ter sido implementados experimentalmente, além disso, podemos ter acesso aos resultados de todos os experimentos feitos que estarão dispostos de acordo com o circuito implementado. A aba \textit{Composer} é essencialmente onde todo o protocolo deve ser inserido e implementado. Ao selecionar essa aba, uma janela será mostrada onde será solicitado ao usuário uma forma de acesso ao \textit{Composer} que basicamente nos permite acessar uma área onde poderemos usar um processador ideal ou real. Enquanto o processador real simula exatamente o que se é esperado no experimento que implementa o circuito onde a computação estará sob o efeito do fenômeno decoerências, o processador ideal não leva isso em conta. Mais a frente discutiremos sobre tais efeitos. A direita da plataforma são encontrados (quando acessamos o processador real) os seguintes "botões" que, agora, passamos a descrever suas respectivas funcionalidades.

\begin{itemize}

\item \textit{Simulate}: Ao construir o circuito nós podemos testar para ver se há algum erro de projeto no circuito. Tais erros possíveis são operações que não podem ser realizadas pelo IBM-QE. É recomendável sempre simular o circuito antes de implementá-lo experimentalmente.

\item \textit{Run}: Uma vez que o circuito passou pelo teste do \textit{Simulate} sem nenhum erro, podemos usar o \textit{Run} para implementar o circuito remotamente em um chip da IBM. Nessa etapa o projeto será submetido ao time da IBM e possivelmente entrará em uma fila de espera até ser implementado.

\item \textit{New}: Nesta opção um novo circuito pode ser criado.

\item \textit{Save} ou \textit{Save as}: Esta opção é destinada ao arquivamento do projeto em qualquer etapa de sua construção.

\item \textit{Results}: Dado a execução de um protocolo, pode-se verificar os resultados nesta opção.

\item \textit{Help}: Fornece informações sobre dúvidas sobre como cada um dos componentes (portas e medidas) podem ser usados e suas formas de atuar sobre os q-bits.

\end{itemize}

Para completar a caracterização da plataforma, precisamos falar do conjunto de portas que podem ser implementadas pelo IBM-QE, uma vez que são esses os elementos fundamentais da plataforma. Na Fig. \ref{circuito} nós podemos ver o conjunto de portas disponíveis na plataforma logo abaixo do circuito. Esse conjunto de porta, composto por portas de Clifford \cite{Samelson:Book} e duas portas não-Clifford que são adicionadas afim de conseguirmos um conjunto compledo de portas universais para computação quântica \cite{Nielsen:Book}. As portas disponíveis são as portas de Pauli ${ X, Y, Z}$, a porta Hadamard ($H$) a porta de fase ($S$ e $S^{\dagger}$), que são portas que podem ser simuladas eficientemente em um computador clássico \cite{Gottesman:98}, mas que quando adicionada das portas não-Clifford ($T$ e $T^{\dagger}$) formam um conjunto de portas nos permite fazer operações não simuladas eficientemente em um computador clássico \cite{Barenco:95}. 

\subsection{Medidas e resultados de uma computação}

Os únicos constituintes do IBM-QE que ainda não mencionamos e que julgamos ser um ponto que deve ser discutido em detalhes é o processo de medida no IBM-QE. A medição é um ponto importante em CQ, principalmente em modelos de CQ baseada em medida, onde todo o processo de computação é feito por meio de uma sequência de medidas sobre os q-bits do sistema \cite{Nielsen:03,Leung:04,Leung:01,Jozsa:05,Gottesman:99}. O IBM-QE tem a disposição do usuário duas maneiras distintas de fazer medidas ao final de um protocolo, cujos símbolos estão dispostos logo abaixo do circuito na Fig. \ref{circuito}.

O primeiro símbolo (logo em seguida da palavra \textit{measure}) é o símbolo de medida na base computacional. Em CQ nós temos a opção de recuperar todos os resultados de uma computação possível em um computador clássico, que por sua vez só faz leituras de $0$ e $1$ que são usados para codificar alguma informação, onde podemos fazer medidas em CQ que são análogas a essas leituras feitas em um computador clássico. Isso graças a possibilidade de fazer medidas na base computacional $\vert 0 \ket$ e $\vert 1 \ket$. A explicação do $Z$ no símbolo da medida na base computacional é que, historicamente, temos primeiro usado de manipulação de spins por meio de campos magnéticos orientados na direção $Z$, onde os estados ortogonais nessa direção foram rotulados como $\vert \uparrow \ket$ e $\vert \downarrow\ket$ associados a estados de spin $+\frac{1}{2}$ e $-\frac{1}{2}$, respectivamente. Na linguagem de CQ esses rótulos podem ser trocados por $\vert \uparrow \ket \rightarrow \vert 0 \ket$ e $\vert \downarrow\ket \rightarrow \vert 1 \ket$. Assim, se estamos simulando computação quântica com spins, como em Ressonância Magnética Nuclear \cite{Cory:00,Roberto:Book}, uma medida na base computacional corresponde a verificarmos se os spins dos constituintes do sistema estão orientados como o campo magnético (orientado na direção $Z$) ou no sentido oposto. 

Como resultado, a medida na base computacional nos informará a \textit{probabilidade} de obter um dado estado como resultado da computação. Por exemplo, considerando um estado como na Eq. (\ref{state2qubits}), uma medida na base computacional nos dará o valor numérico das quantidades $\vert a \vert^2$, $\vert b \vert^2$, $\vert c \vert^2$ e $\vert d \vert^2$. Essa é uma limitação desse tipo de medida, pois não somos capazes de distinguir entre estados como $\vert \pm \ket = ( \vert 0 \ket \pm \vert 1 \ket )/\sqrt{2}$, pois o sinal é indiferente para a probabilidade de medir $\vert 0 \ket$ ou $\vert 1 \ket$. No entanto, para situações onde precisamos caracterizar um estado a fim de saber exatamente qual o estado de saída de uma computação, podemos usar uma medida de Bloch, que é representada pelo símbolo ao lado do símbolo de medida na base computacional.

\begin{figure}[!htb]
\centering
\includegraphics[scale=0.07]{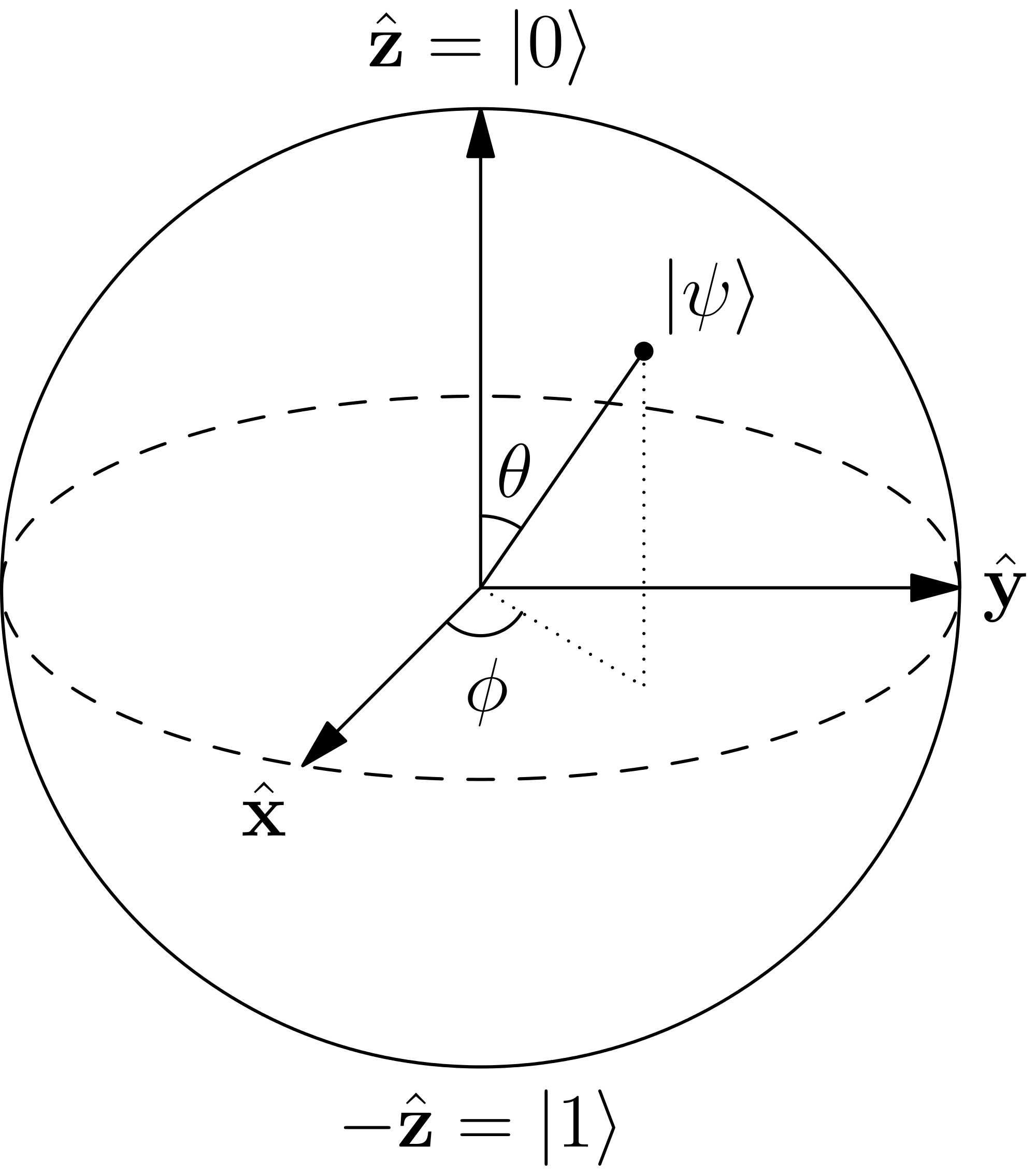}
\caption{Visualização do estado de um q-bit na esfera de Bloch, onde a caracterização do estado se dá por meio do conhecimento das quantidades $(\theta, \phi)$ ou das coordenadas $(x, y, z)$ do vetor $\vert \psi \ket$.}
\label{bloch}
\end{figure}

Para entender a medida de Bloch, precisamos primeiro de recurso para visualizar um estado de um q-bit na esfera de Bloch que pode ser vista na Fig. \ref{bloch}. Diferentemente da medida na base computacional, uma medida de Bloch nos fornece o valor das coordenadas $(x, y, z)$ de um dado estado, que por sua vez podem ser usados para determinar $(\theta, \phi)$ por meio da transformação
\begin{eqnarray}
x = \sin (\theta) \cos (\phi) \text{ , } 
y = \sin (\theta) \sin (\phi) \text{ , } 
z = \cos (\theta) \text{ . } 
\end{eqnarray}

Assim, podemos distinguir com facilidade entre os estados  $\vert \pm \ket$ ou quaisquer outros. Devido à semelhança desse tipo de medida com o conceito de tomografia, em termos específicos da área nós dizemos que essa medida de Bloch é uma \textit{tomografia} de estados \cite{Nielsen:Book}. No entanto, existem situações onde essa medida fracassa, que é quando tentamos fazer a tomografia do estado de dois q-bits emaranhados. O emaranhamento carrega a problemática de que é impossível caracterizar totalmente o estado de um dos q-bits independente do outro, e como o IMB-QE nos fornece estados na esfera de Bloch de um único q-bit, então não temos como usar o processo de tomografia para determinar o estado emaranhado de 2 q-bits. Nesse caso, é mais viável usarmos a medida na base computacional já levando em conta alguma perda de informação sobre quaisquer fases que existam nos estados.

\subsection{Processador Ideal X Processador Real} \label{Deco}

Ao iniciarmos a elaboração de um protocolo (circuito quântico) nós devemos informar para a plataforma do IBM-QE  se desejamos simular nosso circuito em um processador \textit{ideal} ou se optamos por um processador \textit{real}. Esta escolha irá nos direcionar para cenários distintos de mecânica quântica. No processador \textit{ideal} nós realizaremos necessariamente evoluções no sistema sem nenhuma influência de perturbações indesejadas, enquanto que no processador \textit{real} essas perturbações não são ignoradas, nos deixando assim mais próximos de um cenário real de um computador quântico.

Os efeitos indesejados em computação quântica, bem como em qualquer aparato experimental em mecânica quântica, são efeitos de decoerência que estão presentes no processador \textit{real} do IBM-QE. Esses efeitos dependem apenas da forma como nosso sistema interage com o ambiente que o cerca. Não podemos, a partir da aba \textit{User Guide}, obter informações sobre o tipo de decoerência que age sobre o chip de 5 q-bits, mas algumas informações são fornecidas na aba \textit{Composer} quando escolhemos a opção de processador \textit{real}. Uma delas, e no momento mais relevante, é que cada q-bit do chip interage de forma diferente com o ambiente, alguns interagindo mais e outros intragindo menos. Uma das contribuições desse artigo é justamente estudar o tipo de decoerência que age sobre os q-bits e passaremos a discutir isso nesse momento.

\begin{figure}[!htb]
\centering
\subfloat[Circuito]{
\includegraphics[scale=0.4]{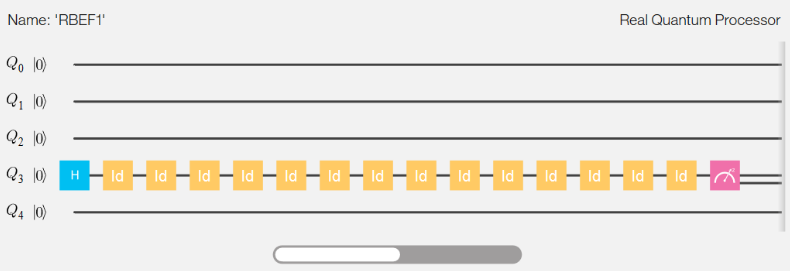}
\label{circuito1}
}
\quad 
\subfloat[Gráfico das amplitudes de probabilidade]{
\includegraphics[scale=0.35]{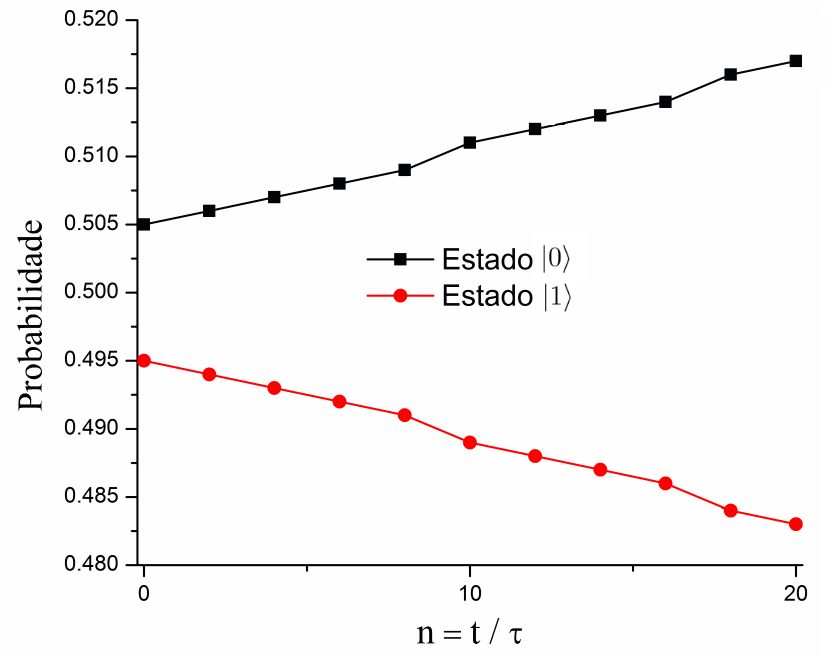}
\label{graph1}
}
\caption{(\ref{circuito1}) Circuito do passo $n=14$ do gráfico presente na Fig. \ref{graph1}. (\ref{graph1}) Gráfico da probabilidade de obter o estado $\vert 0 \ket$ (curva preta) ou $\vert 1 \ket$ (curva vermelha), onde $n$ é o número de portas Identidade (portas que não mudam a computação quando atuam, é um análogo do número 1 no processo de multiplicação de números) utilizadas no circuito e $t$ é o total de evolução com $\tau$ sendo o tempo necessário para implementar cada porta Identidade. Nós usamos o q-bit $3$ pois é o menos robusto a efeitos de decoerência, isto é, sofre mais o efeito do ambiente.}
\label{circuitos1}
\end{figure}

Para estudar os efeitos da decoerência sobre o chip de 5 q-bits do IBM-QE, nós usamos o seguinte protocolo. A primeira parte do circuito é preparar um estado de superposição da forma $ \vert + \ket = \frac{1}{\sqrt{2}} ( \vert 0 \ket + \vert 1 \ket ) $, que pode ser preparado apenas atuando a porta Hadamard no estado de entrada $\vert 0 \ket$. O segundo passo é deixar agora o sistema evoluir apenas sobre a ação dos efeitos de decoerência, onde podemos fazer isso apenas utilizando as portas Identidade (primeira porta de cor laranja na Fig. \ref{circuito} abaixo do circuito) e ao final do processo fazemos uma medida na base computacional (o circuito usado é mostrado na Fig. \ref{circuito1}). Nós utilizamos esse processo e acompanhamos como as amplitudes de probabilidade de obter como resultado o estado $\vert 0 \ket$ e $\vert 1 \ket$, onde o resultado é mostrado no gráfico da Fig. \ref{graph1}.

A partir dos resultados esboçados no gráfico da Fig. \ref{graph1} nós podemos ver que o efeito do ambiente é intensificar a probabilidade de obter o estado de entrada $\vert 0 \ket$, sendo esse um efeito muito parecido com o efeito de emissão espontânea para uma evolução onde $\vert 0 \ket$ é o estado fundamental do sistema. Um exemplo de evoluções desse tipo é por meio de um Hamiltoniano $H = -\hbar \omega \sigma_{z}$, que é o Hamiltoniano de interação de um spin com um campo magnético na direção $Z$, por exemplo. No entanto, esse é apenas um exemplo de evolução que nos permite explicar o que está acontecendo. O que podemos dizer de forma genérica, já que não sabemos explicitamente quais campos atuam sobre os q-bits, é que o efeito do ambiente sobre os q-bits do chip da IBM é o de forçar o sistema para o estado $\vert 0 \ket$.

\subsection{Exemplo: Simulando o circuito do teletransporte} \label{QuantumTeleporteExpe}

\begin{figure}[!htb]
\centering
\includegraphics[scale=0.3]{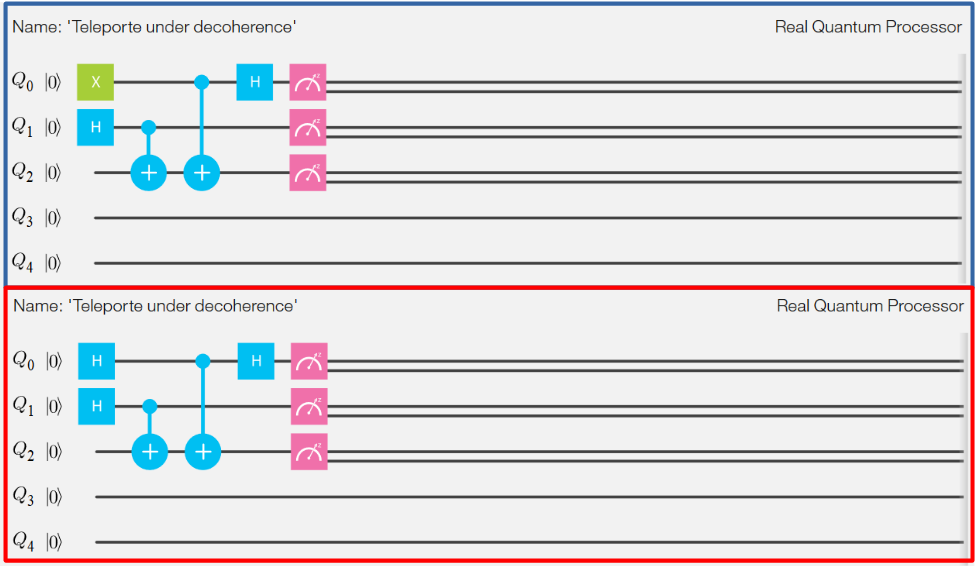}
\caption{Circuitos usados para implementar o teleporte do estado $\vert \psi_{1} \ket = \vert 1 \ket$ (circulado em azul) e $\vert \psi_{2} \ket = \vert + \ket$ (circulado em vermelho). Os estados a serem teleportados são preparados no q-bit $0$ e o recurso é preparado nos q-bits $1$ e $2$, onde o q-bit $2$ está com a Alice e o q-bit $1$ está com o Bob. A porta $X$ (Hadamard) no circuito circulado em azul (vermelho) é usado para preparar o estado a ser teleportado.}
\label{circuitoTele}
\end{figure}

Agora vamos fazer uma aplicação do que temos desenvolvido até agora usando o circuito do teleporte quântico presente na Fig. \ref{BBCprotocol}. Aqui nós ilustraremos nossa aplicação estudando a implementação do teleporte dos estados $\vert \psi_{1} \ket = \vert 1 \ket$ e $\vert \psi_{1} \ket = \vert + \ket$. Os circuitos que implementam o teleporte desses estados são mostrados na Fig. \ref{circuitoTele}.

Nós simulamos os circuitos da Fig. \ref{circuitoTele} com o processador \textit{real}, onde um detalhe deve ser levado em consideração. As portas CNOT que são implementadas no circuito usando um processador real não podem ter qualquer q-bit como o alvo, apenas o q-bit $2$. Como o aparato experimental é baseado no fluxo magnético em q-bits supercondutores, essa é uma restrição do modelo que precisa ser levado em consideração na hora de construir um circuito devido o aparato físico usado.

\begin{figure}[!htb]
\centering
\subfloat[Estado $\vert \psi_{1} \ket = \vert 1 \ket$]{
\includegraphics[scale=0.5]{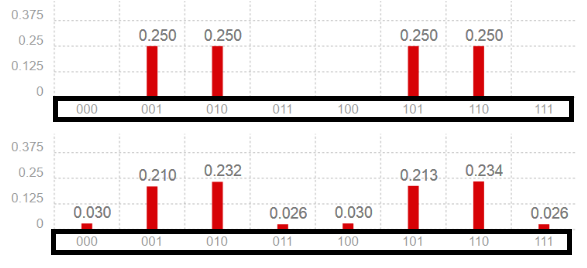}
\label{Estado1}
}
\quad 
\subfloat[Estado $\vert \psi_{2} \ket = \vert + \ket$]{
\includegraphics[scale=0.5]{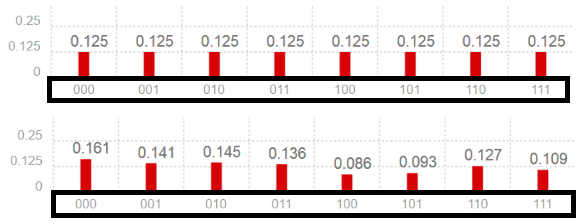}
\label{EstadoMais}
}
\caption{Resultados (distribuição de probabilidade) para o teleporte (\ref{Estado1}) do estado $\vert \psi_{1} \ket = \vert 1 \ket$ e (\ref{EstadoMais}) do estado $\vert \psi_{2} \ket = \vert + \ket$. O gráfico superior em (\ref{Estado1}) e (\ref{EstadoMais}) correspondem ao protocolo realizado em um processador \textit{ideal}, enquanto que o gráfico inferior corresponde a distribuição das probabilidades usando um processador \textit{real}}.
\label{tabelas}
\end{figure}

Quanto ao resultado da implementação do circuito, podemos usar diretamente a Eq. (\ref{ComputatedState}) para mostrar que o estado do sistema antes da medida para o teleporte do estado $\vert \psi_{1} \ket$ deve ser dado por
\begin{equation*}
\vert \phi_{C1} \ket =\frac{1}{2}\left[ \vert
001\ket _{AB}+\vert 010\ket _{AB} - \vert 101\ket _{AB} - \vert 110\ket _{AB} \right] \text{ , } \\
\end{equation*}
onde o estado é teleportado sem necessidade de correção para os resultados $00$ ou $10$, enquanto que este é teleportado a menos de uma correção pela porta $X$ quando os resultados forem $11$ ou $01$. A Fig. \ref{Estado1} mostra os resultados fornecidos pelo IBM-QE quando simulamos o circuito em um processador ideal e real. Nele podemos ver que o estado é teleportado com certeza em, pelo menos, metade das vezes que simulamos o protocolo, com $00$ e $10$ contribuindo com $25 \% $ de acerto cada um.

Para o estado $\vert \psi_{2} \ket = \vert + \ket$ nós temos o seguinte estado final
\begin{equation*}
\vert \phi _{C2}\ket  =\frac{1}{2}\left[ \vert
00+\ket _{AB}+\vert 01+\ket _{AB}+\vert 10-\ket _{AB}-\vert 11-\ket _{AB} \right] \text{ ,} 
\end{equation*}
onde claramente vemos que a probabilidade de sucesso é, novamente sem necessidade de correção, de $50 \% $ correspondente aos resultados $00$ ou $10$. Uma forma equivalente de escrever o estado acima é como
\begin{eqnarray*}
\vert \phi _{C2}\ket  &=&\frac{1}{2}\left[ \left\vert 000\right\rangle _{\text{AB}}+\left\vert 001\right\rangle _{\text{%
AB}}+\left\vert 010\right\rangle _{\text{AB}}+\left\vert 011\right\rangle _{%
\text{AB}} \right] 
\notag \\
&&+\frac{1}{2}\left[ \left\vert 100\right\rangle _{\text{AB}}-\left\vert
101\right\rangle _{\text{AB}}-\left\vert 110\right\rangle _{\text{AB}%
}+\left\vert 111\right\rangle _{\text{AB}}\right] \text{,}
\end{eqnarray*}
onde fica claro a distribuição de probabilidades obtidas na Fig. \ref{EstadoMais} no caso onde simulamos o circuito com um processador \textit{ideal}. Como mencionamos, note que o estado $\vert \phi _{C2}\ket$ é uma superposição de estados da base computacional onde temos explicitamente fases (sinais diferentes) entre os estados, mas nos resultados fornecido pelo IBM-QE presentes na Fig. \ref{EstadoMais} nós não temos essa informação.

\section{Conclusões}

Neste paper nós apresentamos algumas características e funcionalidades do computador quântico da IBM, bem como da plataforma IBM-QE. Ressaltamos suas sutilezas, mostrando que ainda existem limitações no IBM-QE que são intrínsecas do aparato experimental e do modelo de computação utilizado para desenvolvê-lo. Mesmo sem termos conhecimento dos detalhes técnicos acerca dos efeitos da decoerência sobre o chip de 5 q-bits da IBM, nós pudemos identificar um tipo de decoerência existente no sistema analisando um caso particular de evolução. Nossa proposta de estudar fontes de decoerência no IBM-QE pode ser reanalisada para outros estados podem ser construídos, de modo que também outras evoluções podem ser pensadas para tentarmos identificar a existência de outras formas de decoerência no IBM-QE.

Devido ao fato de não termos uma ferramenta didática eficiente para apresentar conceitos como teleporte, robustez de circuitos quânticos contra decoerência, dentre outros conceitos, acreditamos que a plataforma da IBM-QE é uma proposta didática para ser aplicada em disciplinas específicas da área de computação e informação quântica. O teleporte é apenas um exemplo que pode ser discutido, mas existe uma gama de outros protocolos que podem ser trabalhado no intuito de introduzir certos conceitos e problemas em CQ de uma forma mais didática e lúdica.

\section*{Agradecimentos}

Gostaríamos de agradecer a equipe da IBM e aos membros do projeto \textit{IBM Quantum Experience} que, sem dúvidas, tem contribuído não somente para o desenvolvimento desse trabalho mas com toda a comunidade de computação e informação quântica. Em particular, agradeço a Dâmaris Costa Frutuoso, da Universidade Regional do Cariri, por ler o material e sugerir mudanças no mesmo. Agradecemos também ao Conselho Nacional de Desenvolvimento Científico e Tecnológico (CNPq) e ao Instituto Nacional de Ciência e Tecnologia de Informação Quântica (INCT-IQ) pelo suporte financeiro a esse projeto.

\bibliographystyle{ieeetr} 

\end{document}